\documentclass[table]{gtech}
\usepackage[utf8]{inputenc}
\usepackage{amsmath,amssymb}
\usepackage{array}
\usepackage{float}
\usepackage{enumitem}
\usepackage{listings}
\usepackage{algorithm}
\usepackage{algpseudocode}
\algrenewcommand\algorithmicrequire{\textbf{Input:}}
\algrenewcommand\algorithmicensure{\textbf{Output:}}
\usepackage{xurl}
\usepackage{needspace}

\RequirePackage{tgpagella}
\RequirePackage{mathpazo}
\RequirePackage{inconsolata}

\setlist{nosep,leftmargin=*}
\titlespacing*{\section}{0pt}{12pt plus 2pt minus 2pt}{6pt plus 1pt minus 1pt}
\titlespacing*{\subsection}{0pt}{10pt plus 2pt minus 2pt}{4pt plus 1pt minus 1pt}
\titlespacing*{\subsubsection}{0pt}{8pt plus 2pt minus 2pt}{4pt plus 1pt minus 1pt}
\newcommand{\code}[1]{\texttt{#1}}

\renewcommand{\title}[1]{\newcommand{\titlelist}{{\fontsize{17.28}{20.74}\rmfamily\bfseries\selectfont\textcolor{black}{#1}}}}
\title{Reducing Batch Makespan for LLM Agent Workloads through Inference--Environment Coordination}

\author{\mbox{Wei Feng}}
\author{\mbox{Meiling Chen}}
\author{\mbox{Boxiao Chen}}
\author{\mbox{Lu Liu}}
\author{\mbox{Bin Zi}}
\author{\mbox{Libin Ye}}
\author[*]{\mbox{Xingyu Lu}}
\author{\mbox{Lei Xu}}
\author{\mbox{Jun Zhou}}
\affiliation{Ling Team, Ant Group}
\contribution[*]{Corresponding author}
\date{November 11, 2026}
\correspondence{\email{sing.lxy@antgroup.com}}

\abstract{
Efficient batch execution of agent tasks is critical to LLM training and development workflows, including benchmark evaluation, synthetic-data generation, and RL rollouts. Since these tasks alternate between LLM inference and environment interaction, we present coordinated optimizations to reduce batch makespan, the time required to complete all tasks in a batch. For hybrid LLMs, a workload-based resource model guides GPU-memory allocation between key-value (KV) cache and recurrent state and provides a reference sample concurrency to exploit parallelism while limiting recomputation. Trajectory-aware routing and scheduling promote cache reuse while balancing worker load. Under constrained environment capacity, we decouple sample admission from environment execution and introduce Chain-Advancing Action Scheduling (CAAS), which prioritizes long chains using observed progress to shorten the batch tail. Controlled replay combines actual inference with recorded environment responses and durations for ablations of individual optimizations. Live evaluations with Ling-3.0-Flash and Ling-3.1-Flash on the BFCL, SWE-bench Multilingual, WideSearch, and MCP-Atlas benchmarks demonstrate makespan reductions of 10.4\%--65.0\%, while evaluation scores remain aligned. Furthermore, these methods have been successfully applied to training and evaluation workflows for the Ling-3.x model series, accelerating the model-development cycle.
}

\begin{document}
\maketitle
\section{Introduction}

The training and development of agentic LLMs rely on workflows such as benchmark evaluation, synthetic-data generation, and RL rollouts. Repeated across model-development iterations, these workflows often execute agent tasks in batches, with downstream stages awaiting the entire batch. Batch completion time therefore affects development efficiency, motivating our focus on reducing batch makespan, the time required to complete all tasks in a batch, rather than the request-level latency typically optimized in conventional serving, such as time to first token (TTFT) and time per output token (TPOT)~\citep{ref1}.

\begin{figure}[H]
\centering
\includegraphics[width=\linewidth]{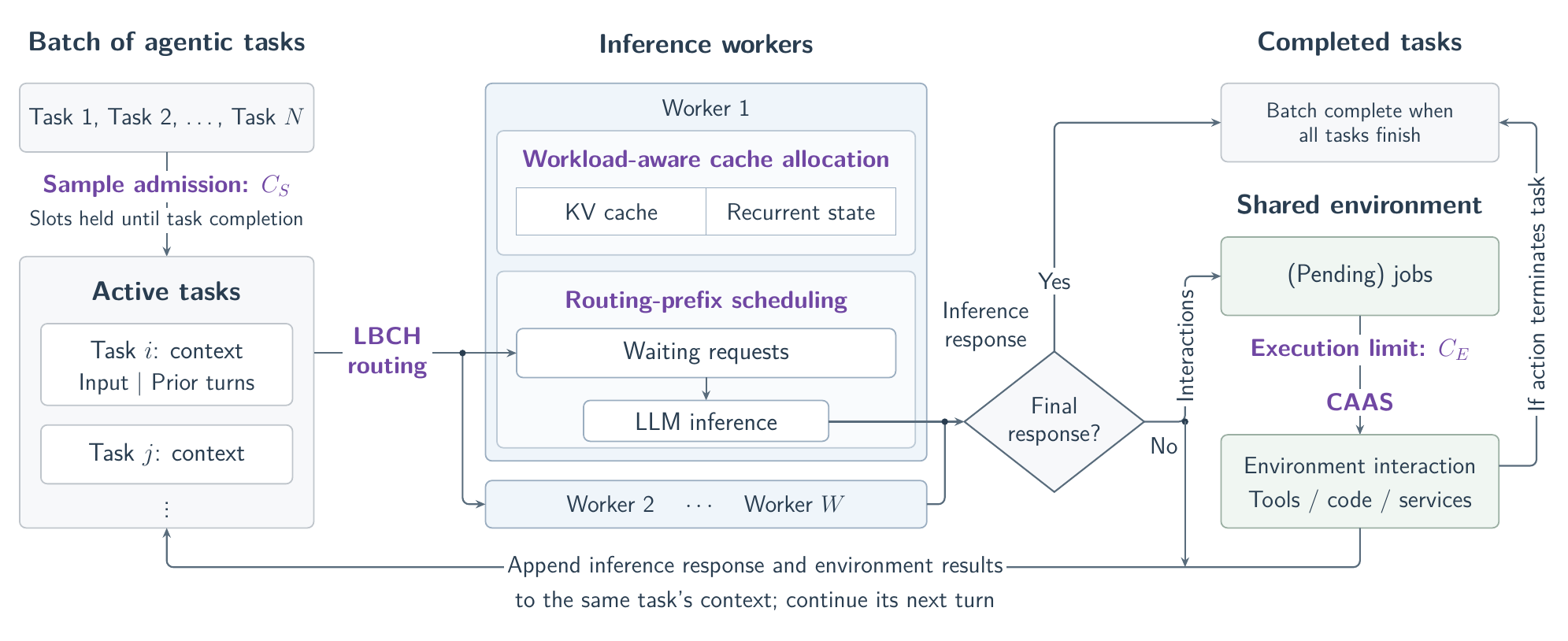}%
\caption{Illustration of batched execution of multi-turn LLM agent tasks. Active tasks are those admitted but not yet completed, including tasks waiting for inference or environment interaction. Each maintains its own context while sharing inference and environment-execution resources. Purple labels indicate where our methods are applied. Capacity separation allows the sample-admission limit \(C_S\) and environment-execution limit \(C_E\) to be configured independently. Our workload-aware capacity model guides the GPU memory allocation between KV cache and recurrent state at deployment; the allocation remains unchanged during batch execution.}
\label{fig:system-overview}
\end{figure}

We refer to an agentic task instance as a sample and its complete execution as a trajectory. Figure~\ref{fig:system-overview} illustrates how a batch of such trajectories is executed, each alternating between LLM inference and environment interaction. Tasks are admitted under a sample-concurrency limit, and their inference requests are routed to workers and ordered by local schedulers. Actions produced by the model are queued as environment jobs and dispatched under an execution limit. Model responses and environment results are appended to the task's context before the next inference turn. A final response or terminal environment action completes the task and releases its admission slot. The batch finishes when all tasks complete.

\enlargethispage{\baselineskip}
Successive requests within a trajectory share growing input prefixes, creating opportunities for cache reuse. Hybrid models such as Ling-3.0-Flash~\citep{ref50}, Qwen3.8-Max~\citep{ref56}, and Kimi K3~\citep{ref57} require both KV cache and recurrent state for cross-turn reuse. For a given model, KV-cache memory grows with the number of cached tokens, whereas each recurrent state has a fixed size. Thus, longer contexts increase KV-cache demand per trajectory, while admitting more concurrent trajectories requires retaining more recurrent states as well as their associated KV cache. SGLang manages these states in separate pools within a shared cache-memory budget. If a default or empirically chosen allocation ratio does not accurately reflect the target workload's cache-demand characteristics, one pool can be exhausted while substantial capacity remains unused in the other. This imbalance limits the number of trajectories whose caches can be retained together. Our workload-aware capacity model quantifies the target workload's recurrent-state and KV-cache requirements to balance memory allocation between the two pools at deployment and provide a reference sample concurrency, enabling greater parallelism while preserving cross-turn cache reuse. The resulting allocation remains fixed during inference serving.

Even with suitable allocation, reuse depends on where and when requests execute. With multiple inference workers, keeping a trajectory's requests on the same worker favors reuse, but strict affinity can concentrate load and lengthen queues. Even on a worker holding reusable state, a request can wait long enough for that state to be evicted. In addition, model responses and environment results accumulate across turns, potentially producing very long agent contexts. Identifying cache-reuse opportunities through token-prefix matching can then incur substantial computational overhead. To reduce this overhead, we use a shared trajectory label to identify requests from the same trajectory at low cost, giving the router an affinity key and letting the worker screen waiting requests before querying the prefix cache for scheduling priority. We implement this approach through two complementary methods. \textit{Load-bounded consistent hashing (LBCH)} routing uses the labels to preserve cross-turn worker affinity while accounting for worker load. \textit{Routing-prefix (RP)} scheduling uses them to prioritize requests from recently served trajectories and promote timely cache reuse at low scheduling cost.

Environment resources may also be constrained by CPU budgets or external-service quotas. A single sample-concurrency limit, as used in the agent implementations of MCP-Atlas and \mbox{\(\tau^3\)-bench}~\citep{ref51,ref52}, couples the two stages. When the environment supports less concurrency than the inference service, raising this shared limit to keep inference busy can overload environment execution. Conversely, lowering it to protect the environment can leave inference workers without ready requests while active samples await environment-interaction results. We therefore decouple sample admission from environment execution, allowing more samples to remain active while an independent execution limit protects the environment. This improves stage overlap but also makes environment-side queuing more likely as more admitted samples submit jobs to the limited execution slots. Their dispatch order determines when each trajectory can resume inference. Repeated waits can defer a long trajectory's later stages until most other samples have finished, leaving few ready inference requests and prolonging the batch tail. When multiple environment jobs await execution, Chain-Advancing Action Scheduling (CAAS) uses observed progress to prioritize those from potentially long trajectories. This aims to expose their subsequent stages sooner and shorten the batch tail without requiring knowledge of remaining trajectory lengths.

Our contributions are summarized as follows.
\begin{enumerate}
\item \textbf{Workload-based and trajectory-aware inference optimization.} A hybrid-cache capacity model guides allocation and sample concurrency, complemented by LBCH routing for affinity--load balance and routing-prefix scheduling for timely cross-turn reuse.
\item \textbf{Inference--environment coordination.} Independent sample and environment concurrency controls improve stage overlap, while CAAS effectively advances long execution chains to shorten the batch tail.
\item \textbf{Controlled replay and live end-to-end validation.} Trace-driven replay combines actual inference with recorded environment responses and durations for controlled comparisons and ablations. Live evaluations with Ling-3.0-Flash and Ling-3.1-Flash across four well-known benchmarks demonstrate substantial reductions in batch makespan, while evaluation scores remain aligned (Section~\ref{sec:end-to-end-validation}). The proposed methods are also applied in Ling-3.x model development, where batched agent workloads such as benchmark evaluation, synthetic-data generation, and RL rollouts play a key role. By accelerating these workloads, the methods substantially improve overall development efficiency.
\end{enumerate}

The remainder of this paper presents the methods and experimental evaluations supporting these contributions. Section~\ref{sec:related-work} reviews related work. Sections~\ref{sec:inference} and~\ref{sec:coordination} present inference-side optimizations and inference--environment coordination, respectively. Section~\ref{sec:evaluation} evaluates the proposed methods through controlled replay and live execution. Section~\ref{sec:discussion} discusses future directions, and Section~\ref{sec:conclusions} concludes the paper.

\section{Related Work}\label{sec:related-work}
\textbf{Prefix caching.} Prefix caching avoids repeated computation over shared prompt prefixes. Serving systems support prefix reuse through mechanisms such as SGLang's RadixAttention~\citep{ref2} and vLLM's automatic prefix caching~\citep{ref49}. Mooncake, Tutti, and CacheFlow extend cache management across distributed or multi-tier settings~\citep{ref20,ref36,ref37}. Li et al.~\citeyearpar{ref47} survey these broader cache-management techniques. BatchLLM exploits globally available prefix structure and throughput-oriented batching for large inference batches~\citep{ref-batchllm}, whereas we target cross-turn reuse within agent trajectories.

\textbf{Cache management for hybrid LLMs.} Cache reuse in hybrid LLMs requires retaining both KV cache and recurrent state. AVMP manages separate KV-cache and state-space-model (SSM) state pools through a unified virtual page table~\citep{ref38}. It can transfer capacity between pools at runtime when allocation fails in one pool and the other has enough spare capacity. Marconi selects recurrent states for caching based on their reuse potential and guides eviction using recency and computation saved relative to the combined footprint of the KV cache and recurrent state~\citep{ref58}. Our workload-aware capacity model uses trajectory profiles to select a static partition of the cache budget and a reference sample concurrency before serving, with the goal of increasing joint trajectory-cache residency.

\textbf{Cache-aware routing and scheduling.} Preble uses global prefix-cache information and computation estimates to route requests across workers~\citep{ref26}. Its local scheduler groups requests by the fraction of cached prompt tokens and selects requests from each group to form a batch, assigning larger shares to higher-priority groups. DualMap hashes a prompt prefix to two candidate workers and uses a time-to-first-token (TTFT) service-level objective to balance affinity and load~\citep{ref28}. PEEK maintains an incremental radix tree over pending requests to identify shared prefixes even before they are cached~\citep{ref29}. It uses these relationships to schedule prefix-sharing requests and protect cache entries needed by queued requests. These methods use prompt-prefix information to guide routing or scheduling. In contrast, our LBCH routing and RP scheduling use trajectory labels to identify and exploit potential cache-reuse opportunities.

\textbf{Agent-aware inference scheduling.} Autellix schedules LLM calls using observed program-level attained service and uses program locality in cross-engine routing~\citep{ref32}. Helium represents workflows as query plans and optimizes workflow makespan through rewriting, caching, and cost-based scheduling~\citep{ref30}. SAGA uses predicted workflow structure for cache management and workflow-level scheduling~\citep{ref31}. ConServe uses observable conversation state to place the first-turn prefill and maintains decoder affinity across turns~\citep{ref34}. RP links successive turns of an agent trajectory through shared labels and recent serving activity, without tracking program-level attained service or requiring workflow graphs.

\textbf{Environment-aware serving and action scheduling.} SAGA uses expected tool-call durations to guide cache retention~\citep{ref31}, while MORI uses relative tool-call idleness to guide cache placement across GPU and CPU memory~\citep{ref33}. Both use environment behavior to inform inference-side cache decisions. ARL-Tangram directly schedules external actions, using first-come, first-served (FCFS) to select candidates and elastic resource allocation to reduce the sum of action completion times~\citep{ref55}. We instead order pending environment jobs under a fixed execution limit with separate sample admission. CAAS uses observed trajectory progress to advance long execution chains and shorten the batch tail without changing individual job service times.

\section{Optimizing Inference Efficiency}
\label{sec:inference}

This section presents our methods for improving inference efficiency, which are implemented in SGLang~\citep{ref2}. We first examine cache allocation for hybrid models, then present cross-worker routing and in-worker scheduling to improve cross-turn reuse.

\subsection{Workload-Aware Hybrid-Cache Allocation}\label{sec:allocation}

Hybrid models such as Ling-3.0-Flash~\citep{ref50} combine Multi-head Latent Attention (MLA) with Kimi Delta Attention (KDA), requiring KV cache and recurrent state, respectively. Each inference worker maintains separate pools for these states within a shared cache-memory budget. Cache reuse requires both types of state for the reused prefix to remain available, so the more constrained pool limits how many trajectories can retain reusable state.

KV-cache demand grows with cached context length, whereas recurrent state occupies a fixed number of slots per request. As a trajectory extends its prefix, the latest reusable recurrent state summarizes that prefix and supports subsequent computation without retaining states from earlier turns. We therefore retain one recurrent state per trajectory, keeping its recurrent-state memory requirement fixed as the number of turns increases.

At low sample concurrency, performance is relatively insensitive to the cache ratio when both pools have sufficient capacity. Higher concurrency can shorten makespan, but also increases cache demand, so an imbalanced allocation can cause eviction and recomputation that offset parallelism gains. We therefore use a workload-aware capacity model to guide cache allocation and sample concurrency, aiming to sustain high cross-turn reuse while increasing parallelism.

Model development often involves repeated execution of batched agent workloads, including benchmark evaluation, synthetic-data generation, and RL rollouts, across successive checkpoints. For similar task distributions, the mean context length often changes little between nearby checkpoints, allowing historical trajectory profiles to provide reusable estimates for guiding cache allocation and sample concurrency.

\subsubsection{Joint Capacity Model}

Let \(M\) be the GPU memory budget available to the KV-cache and recurrent-state pools after model weights and non-cache runtime reservations. Let \(M_{\mathrm{kv}}\) and \(M_{\mathrm{state}}\) denote the memory allocated to the two pools, respectively, and define \(r=M_{\mathrm{state}}/M_{\mathrm{kv}}\). The resulting partition is

\begin{equation}
M_{\mathrm{state}}(r) = M \frac{r}{1+r}, \quad
M_{\mathrm{kv}}(r) = M \frac{1}{1+r}.
\label{eq:memory-partition}
\end{equation}

For workload \(\mathcal D\), we combine historical trajectory profiles with model and runtime parameters to estimate the per-trajectory cache footprints as

\begin{equation}
F_{\mathrm{kv}}(\mathcal D) = L_{\mathrm{kv}}(\mathcal D)c_{\mathrm{kv}}, \quad
F_{\mathrm{state}} = S_{\mathrm{state}} c_{\mathrm{state}},
\label{eq:trajectory-cache-footprint}
\end{equation}
where \(L_{\mathrm{kv}}(\mathcal D)\) is the average number of KV-cache entries required by a complete trajectory in \(\mathcal D\), measured at its final inference turn, \(S_{\mathrm{state}}\) is the number of recurrent-state slots required by each inference request under the deployment configuration, and \(c_{\mathrm{kv}}\) and \(c_{\mathrm{state}}\) are their respective unit memory costs. The quantities \(S_{\mathrm{state}}\), \(c_{\mathrm{kv}}\), and \(c_{\mathrm{state}}\) are fixed for a given model and runtime configuration. Because the accumulated context generally grows across turns, the final-turn KV-cache allocation provides a conservative workload estimate, although it is not treated as a strict upper bound.

Based on the pool allocations in Equation~\ref{eq:memory-partition} and the workload-derived cache footprints in Equation~\ref{eq:trajectory-cache-footprint}, we approximate the number of trajectories whose cache state can be retained in each pool simultaneously using the following continuous estimates:

\begin{equation}
C_{\mathrm{kv}}(r;\mathcal D) = \frac{M_{\mathrm{kv}}(r)}{F_{\mathrm{kv}}(\mathcal D)}, \quad
C_{\mathrm{state}}(r) = \frac{M_{\mathrm{state}}(r)}{F_{\mathrm{state}}}.
\label{eq:resource-residency-estimates}
\end{equation}
Under this per-trajectory footprint approximation, we estimate the joint residency capacity as the minimum of the two pool-specific estimates:

\begin{equation}
C(r;\mathcal D)
=\min\!\left\{C_{\mathrm{kv}}(r;\mathcal D),C_{\mathrm{state}}(r)\right\}.
\label{eq:resident-trajectory-estimate}
\end{equation}

The joint capacity \(C(r;\mathcal D)\) estimates how many trajectories can simultaneously retain both KV cache and recurrent state on a worker at ratio \(r\), providing a reference for sample concurrency. The analytical model treats the available cache memory as continuously divisible and abstracts away allocator-level details such as recurrent-state slot granularity, KV-page granularity, non-contiguous memory placement, and runtime fragmentation.

\subsubsection{Capacity-Balancing Estimator}

Under the continuous capacity model, joint residency is limited by the pool that can support fewer trajectories. Reallocating memory from the other pool to this limiting pool increases the joint capacity until both pools support the same number of trajectories. Setting \(C_{\mathrm{kv}}(r;\mathcal D)=C_{\mathrm{state}}(r)\) gives

\begin{equation}
\frac{M_{\mathrm{kv}}(r)}{F_{\mathrm{kv}}(\mathcal D)}
=
\frac{M_{\mathrm{state}}(r)}{F_{\mathrm{state}}},
\label{eq:capacity-balance-condition}
\end{equation}
which yields the balancing target

\begin{equation}
r^{\mathrm{bal}}(\mathcal D)
=
\frac{F_{\mathrm{state}}}{F_{\mathrm{kv}}(\mathcal D)}
=
\frac{S_{\mathrm{state}}c_{\mathrm{state}}}
     {L_{\mathrm{kv}}(\mathcal D)c_{\mathrm{kv}}}.
\label{eq:balanced-allocation-ratio}
\end{equation}

At this balancing ratio, the two pools support the same number of trajectories, and the joint capacity simplifies to
\begin{equation}
C(r^{\mathrm{bal}}(\mathcal D);\mathcal D)
=\frac{M}{F_{\mathrm{kv}}(\mathcal D)+F_{\mathrm{state}}}.
\label{eq:balanced-residency-capacity}
\end{equation}

This balancing ratio may not be optimal in practice, since the workload-wide mean KV-cache footprint may not reflect the cache demand of the trajectories active at a given time. Optional local calibration can identify a better-performing nearby ratio. We use \(\hat r\) to denote the selected cache ratio, which can be either the model-derived value or a locally calibrated value. For deployment, the continuous capacity estimate \(C(\hat r;\mathcal D)\) is rounded down to obtain the integer reference concurrency \(\lfloor C(\hat r;\mathcal D)\rfloor\) per worker.
For \(W\) homogeneous workers, we use the aggregate reference concurrency
\begin{equation}
C_I^\star(\hat r;\mathcal D)
=
W\left\lfloor C(\hat r;\mathcal D)\right\rfloor.
\label{eq:cache-efficient-sample-concurrency}
\end{equation}
The selected ratio \(\hat r\) stays fixed during inference serving. The reference concurrency does not impose a hard limit, since timely cache reuse may support higher concurrency.
The results in Section~\ref{sec:cache-allocation-evaluation} show that the model-derived cache ratios and reference concurrency settings provide effective operating points for the evaluated workloads.

\subsection{Cross-Worker Trajectory-Aware Routing}

We consider a distributed inference deployment comprising a gateway and multiple homogeneous workers. Each worker is a model-serving replica that independently schedules its assigned inference requests and maintains its own device-local KV-cache and recurrent-state pools. The gateway routes each inference turn to one of these workers. We focus on worker-local cache reuse without cross-worker cache transfer, so a later turn can reuse its trajectory cache only on the worker that currently holds the corresponding cached state.

Routing decisions based solely on load balancing may send successive turns of the same trajectory to different workers, missing opportunities for cache reuse. Conversely, prioritizing trajectory affinity without considering worker load can concentrate requests on a few workers. Requests may then queue at these workers while others remain idle, reducing overall inference efficiency. We adapt consistent hashing with bounded loads \citep{ref-mirrokni} to per-turn inference routing. LBCH uses a shared trajectory label for affinity and current in-flight request counts for load control. This routing method requires every inference request in trajectory \(k\) to carry the same trajectory label \(\kappa_k\). Let \(\mathcal W\) denote the worker set and \(W=|\mathcal W|\) its size. These workers are organized on a hash ring \(\mathcal R\), and \(\kappa_k\) is hashed to a position on \(\mathcal R\). The first worker clockwise from that position is defined as the home worker \(h(\kappa_k)\) of trajectory \(k\). Let \(L(w)\) denote the number of in-flight inference requests assigned to worker \(w\) that are either queued or running, and let \(L_{\mathrm{tot}}=\sum_{w\in\mathcal W}L(w)\). A worker is eligible when

\begin{equation}
L(w)
\leq
\frac{L_{\mathrm{tot}}+1}{W}\alpha,
\label{eq:bounded-load-condition}
\end{equation}
where the \(+1\) term accounts for the request currently being assigned, and \(\alpha \ge 1\) is the load-bound factor. Values above \(1\) permit controlled load skew in favor of trajectory affinity, whereas \(\alpha=1\) gives the tightest eligibility threshold. The router selects \(h(\kappa_k)\) if it is eligible. Otherwise, it traverses the ring clockwise with the successor operator \(\operatorname{succ}_{\mathcal R}(w)\), which returns the next worker clockwise from \(w\) on \(\mathcal R\), and selects the first eligible successor. If no successor is eligible, it falls back to the home worker. Therefore, the resulting procedure, summarized in Algorithm~\ref{alg:trajectory-aware-routing}, relies only on trajectory labels and worker loads, without inspecting prompt text or predicting cache contents. The experiments in Section~\ref{sec:cross-worker-routing-evaluation} demonstrate that LBCH reduces request-load imbalance across workers while maintaining high cache reuse.

\begin{algorithm}[!htbp]
\caption{Trajectory-aware routing with load-bounded consistent hashing (LBCH)}
\label{alg:trajectory-aware-routing}
\begin{algorithmic}[1]
\Require trajectory label \(\kappa_k\), hash ring \(\mathcal R\), loads \(\{L(w)\}_{w\in\mathcal W}\), load-bound factor \(\alpha\)
\Ensure selected worker
\State \(w_0 \gets h(\kappa_k)\)
\State \(\tau \gets \frac{L_{\mathrm{tot}}+1}{W}\alpha\)
\If{\(L(w_0)\leq \tau\)}
  \State \Return \(w_0\)
\EndIf
\State \(w \gets w_0\)
\Repeat
  \State \(w \gets \operatorname{succ}_{\mathcal R}(w)\)
  \If{\(w\neq w_0\) \textbf{and} \(L(w)\leq\tau\)}
    \State \Return \(w\)
  \EndIf
\Until{\(w=w_0\)}
\State \Return \(w_0\)
\end{algorithmic}
\end{algorithm}

\subsection{In-Worker Routing-Prefix Scheduling}

Worker affinity alone cannot guarantee cache reuse. At high sample concurrency, a request may wait in the worker's queue long enough for its trajectory's cached state to be evicted. When the request is eventually served, the worker must recompute the affected prefix. Routing-prefix prioritizes queued requests from recently served trajectories to promote timely reuse while limiting prefix-matching overhead.

For each worker, let \(\mathcal Q=(q_1,\ldots,q_n)\) be its waiting queue, \(\mathcal A\) its trajectory-label set with capacity \(C_{\mathcal A}\), and \(\mathcal C\) its radix prefix cache. The trajectory-label set is updated using a least recently used (LRU) rule. Algorithm~\ref{alg:routing-prefix-scheduling} summarizes the policy in three stages: scoring the waiting requests, building the next batch, and updating \(\mathcal A\). Each request \(q\) has a trajectory label \(\kappa(q)\), a cached prefix-hit length \(\ell(q)\), and a queue-entry time \(t_{\mathrm{queue}}(q)\). It also carries a binary prefix-query flag \(\delta_q\in\{0,1\}\), initialized to \(1\) when the request is created. During scoring, requests whose labels belong to \(\mathcal A\) form the first tier. The operator \(\operatorname{PrefixHit}(\mathcal C,q)\) returns the maximum prefix-match length for \(q\) in \(\mathcal C\), measured in tokens. For each first-tier request with \(\delta_q=1\), the scheduler queries \(\mathcal C\) via \(\operatorname{PrefixHit}(\mathcal C,q)\), records the resulting value as \(\ell(q)\), and sets \(\delta_q\) to \(0\); when \(\delta_q=0\), the recorded \(\ell(q)\) is reused. For requests with \(\kappa(q)\notin\mathcal A\), the scheduler skips the prefix-cache query. The priority key is defined as
\[
p(q)=
\begin{cases}
(0,-\ell(q),t_{\mathrm{queue}}(q)), & \kappa(q)\in\mathcal A,\\
(1,0,t_{\mathrm{queue}}(q)), & \kappa(q)\notin\mathcal A.
\end{cases}
\]
The operator \(\operatorname{Sort}^{\uparrow}_{\mathrm{lex}}(\mathcal Q,p)\) orders the requests in \(\mathcal Q\) by ascending lexicographic order of their priority keys \(p(q)\). In this ordering, the first component places requests with labels in \(\mathcal A\) ahead of the others, the second ranks those requests by decreasing prefix-hit length, and the third preserves queue-entry order when the preceding components are tied.

\begin{algorithm}[!htbp]
\caption{Routing-prefix policy}
\label{alg:routing-prefix-scheduling}
\begin{algorithmic}[1]
\Require waiting queue \(\mathcal Q\), label set \(\mathcal A\), capacity \(C_{\mathcal A}\), radix prefix cache \(\mathcal C\)
\Ensure admitted batch \(\mathcal B\), remaining queue \(\mathcal Q\), updated label set \(\mathcal A\)
\ForAll{\(q\in\mathcal Q\)}
  \If{\(\kappa(q)\in\mathcal A\)}
    \If{\(\delta_q=1\)}
      \State \(\ell(q)\gets\operatorname{PrefixHit}(\mathcal C,q)\)
      \State \(\delta_q\gets 0\)
    \EndIf
    \State \(p(q)\gets(0,-\ell(q),t_{\mathrm{queue}}(q))\)
  \Else
    \State \(p(q)\gets(1,0,t_{\mathrm{queue}}(q))\)
  \EndIf
\EndFor
\State \(\mathcal Q\gets\operatorname{Sort}^{\uparrow}_{\mathrm{lex}}(\mathcal Q,p)\)
\State \((\mathcal B,\mathcal Q)\gets\operatorname{BuildBatch}(\mathcal Q)\)
\ForAll{\(q\in\mathcal B\)}
  \State \(\mathcal A\gets\operatorname{LRUUpdate}(\mathcal A,\kappa(q),C_{\mathcal A})\)
\EndFor
\State \Return \(\mathcal B,\mathcal Q,\mathcal A\)
\end{algorithmic}
\end{algorithm}

After this reordering, the existing batch builder scans \(\mathcal Q\) in order and admits requests subject to the current resource limits. This operation is denoted by \((\mathcal B,\mathcal Q)\gets\operatorname{BuildBatch}(\mathcal Q)\), where \(\mathcal B\) is the admitted batch and \(\mathcal Q\) is the remaining queue. For admitted requests only, \(\mathcal A\) is updated using their labels according to
\[
\operatorname{LRUUpdate}(\mathcal A,\kappa,C_{\mathcal A})
=
\operatorname{tail}_{C_{\mathcal A}}
\bigl((\mathcal A\setminus\{\kappa\})\mathbin{\|}\langle\kappa\rangle\bigr),
\]
where \(\mathbin{\|}\) appends \(\kappa\) at the most-recent position and \(\operatorname{tail}_{C_{\mathcal A}}\) retains the \(C_{\mathcal A}\) most recent labels.

The label set \(\mathcal A\) records recent batch admissions, not GPU cache residency. Its LRU updates are independent of GPU cache eviction, so labels and cached state can become inconsistent in two ways. First, a label may remain in \(\mathcal A\) after the corresponding cached state has been evicted. If eviction occurs before the prefix lookup, the resulting hit length may be short or zero, potentially lowering the request's rank within the first tier. If eviction occurs after the lookup, the stored hit length may become stale because subsequent priority calculations reuse it without another lookup. Second, a label may be removed from \(\mathcal A\) while corresponding cached state remains on the worker: removing a label does not evict that state. A request with this label skips the priority-related prefix lookup and enters the second tier, potentially missing an opportunity to prioritize reuse of the remaining cached state.

The capacity \(C_{\mathcal A}\) therefore controls label coverage, not GPU cache allocation. Larger sets offer more reuse opportunities at higher prefix-query cost, while smaller sets reduce this cost but may miss reuse opportunities. We use \(C_{\mathcal A}=\lfloor C(\hat r;\mathcal D)\rfloor\) as a model-informed default, matching the set capacity to the estimated number of trajectories whose KV cache and recurrent state one worker can retain under the selected cache ratio. Section~\ref{sec:in-worker-scheduling} evaluates whether this RP design preserves cache reuse and limits priority-calculation cost under cache pressure.

\subsection{Integration with SGLang}
\label{sec:sglang-integration}

At deployment, we apply the selected cache ratio $\hat r$ through SGLang's native \code{mamba\_full\_memory\_ratio} parameter. Since each trajectory requires only its latest reusable recurrent state, as explained in Section~\ref{sec:allocation}, we use SGLang's native path-based Mamba-state eviction with \code{mamba\_max\_states\_per\_path=1} to prune older states as the trajectory advances. We add LBCH as a routing policy in SGLang's router and RP as a scheduling policy in each worker. Both policies use the same trajectory label carried as request metadata. LBCH selects a worker using this label and in-flight request counts. RP reorders the waiting queue while reusing SGLang's existing prefix-cache lookup and resource-constrained batch builder. Sample admission remains controlled by the agent execution layer, as described in Section~\ref{sec:coordination}.

\section{Coordinating Inference and Environment Interaction}
\label{sec:coordination}

In the agent execution layer, we decouple sample admission from environment execution by setting their concurrency limits independently, and use CAAS to dispatch pending environment-interaction jobs. Trajectory $k$ has $N_k$ inference stages $I_{k,n}$, with $E_{k,n}$ denoting the corresponding environment-interaction stage, when present. Execution follows
\begin{equation}
I_{k,1}\rightarrow E_{k,1}\rightarrow I_{k,2}
\rightarrow \cdots \rightarrow I_{k,N_k}
\bigl(\rightarrow E_{k,N_k}\bigr)
\label{eq:alternating-chain}
\end{equation}
and completes after the final inference stage or, if present, the corresponding environment stage. Completing a nonterminal environment stage releases the next inference turn, so each trajectory has at most one outstanding environment stage.

\subsection{Separating Sample Admission from Execution Capacity}

Although all samples are eligible at batch start, admitting them all at once can overload inference and environment services. The sample-admission limit $C_S$ bounds the number of active samples. Each sample holds a slot from admission to completion, including while waiting for inference or environment interaction. Waiting samples are admitted in submission order as slots become available. Each environment-interaction stage is modeled as a non-preemptive job comprising one or more operations. The environment-execution capacity $C_E$ limits the number of concurrently running jobs.

In practice, agent-evaluation harnesses such as MCP-Atlas~\citep{ref51} and $\tau^3$-bench~\citep{ref52} commonly use a single task-level concurrency limit. We model this as a unified concurrency gate that constrains sample admission by both inference and environment capacity, giving
\begin{equation}
C_S^{\mathrm{unified}}
=\min\!\left \{ C_I^\star(\hat r;\mathcal D),C_E\right \},
\label{eq:unified-admission}
\end{equation}
where $C_I^\star$ is the inference-side reference concurrency from Section~\ref{sec:allocation}. When $C_E<C_I^\star$, this coupling can underutilize GPUs while samples await environment results. We instead default to
\begin{equation}
C_S^{\mathrm{decoupled}}
=C_I^\star(\hat r;\mathcal D)
\label{eq:decoupled-admission}
\end{equation}
and independently cap running environment-interaction jobs at $C_E$. Pending jobs retain sample slots but no environment-execution slots. Figure~\ref{fig:coordination-capacity} demonstrates the benefits of improved\linebreak inference--environment overlap.

\subsection{Chain-Advancing Action Scheduling}

Capacity separation allows more active trajectories under the same environment-execution limit, which can increase queueing for environment jobs. Their dispatch order determines which trajectory resumes inference next. In practice, a few long trajectories often prolong batch completion after most samples have finished, further reducing GPU utilization. CAAS therefore aims to identify and advance potentially long trajectories earlier through environment-job scheduling to shorten batch makespan.

We consider a homogeneous execution pool where any pending job can use any available slot. When a slot becomes available, CAAS selects the next job without changing the agent's actions or job service times. For job $j$, let $a_j$ be its slot-request time, $n_j$ the number of completed inference turns in its trajectory, including the turn that generated job $j$, and $\tau_m$ the observed duration of its $m$th inference turn. With hyperparameters $B>0$ and $\Delta_I>0$, we define
\begin{equation}
g_j=\left\lfloor \frac{a_j}{B}\right\rfloor,
\label{eq:cats-time-band}
\end{equation}
\begin{equation}
\bar{\tau}_j=\frac{1}{n_j}\sum_{m=1}^{n_j}\tau_m,
\label{eq:cats-average-inference}
\end{equation}
\begin{equation}
c_j=
\left\lceil \bar{\tau}_j/\Delta_I\right\rceil.
\label{eq:cats-inference-cost}
\end{equation}
Here $g_j$ groups jobs by arrival time, while $c_j$ discretizes their observed mean inference duration. When job $j$ joins the pending queue, CAAS assigns it the priority key
\begin{equation}
p_E(j)=\bigl(g_j,-n_j,c_j,e_j\bigr),
\label{eq:cats-priority}
\end{equation}
where $e_j$ is a globally increasing entry sequence. Each job's key remains fixed while it waits, and pending jobs are sorted in ascending lexicographic order. The ordering first protects jobs in earlier arrival bands through $g_j$, then prioritizes deeper revealed chains through $-n_j$. Within the same band and depth, $c_j$ provides a coarse tie breaker favoring a shorter estimated inference return, with $e_j$ resolving remaining ties in FCFS order. A smaller $B$ favors arrival order, whereas a larger $B$ gives chain depth more influence. Duration bucketing reduces noise in the inference-duration tie breaker.

CAAS makes online dispatch decisions using only trajectory state observed in the current run, without requiring sample-specific historical profiles or predictions of total or remaining execution time. With future trajectory lengths unknown, CAAS uses completed inference turns as a proxy for potentially long trajectories and prioritizes deeper chains within each arrival band. This allows subsequent stages to be revealed and executed sooner and reduces the risk that long chains are left to dominate batch completion. Arrival bands prevent returning trajectories from repeatedly overtaking shallower jobs waiting in earlier bands, as can happen under depth-only ordering. Section~\ref{sec:tool-coordination-evaluation} shows that CAAS reduces batch tail and makespan over FCFS after capacity separation.

\section{Experimental Evaluation}
\label{sec:evaluation}

Our evaluation examines how the proposed optimizations reduce batch makespan, both individually and in combination. Section~\ref{sec:replay-framework} first describes the replay framework and experimental setup. Using controlled replay, we evaluate how cache allocation, sample concurrency, scheduling, and routing affect inference efficiency (Section~\ref{sec:inference-evaluation}), and isolate the effects of capacity separation and CAAS under constrained environment capacity (Section~\ref{sec:tool-coordination-evaluation}). Live experiments then evaluate their combined effects across models of different sizes (Section~\ref{sec:end-to-end-validation}).

\subsection{Controlled Replay Framework and Setup}
\label{sec:replay-framework}

Real-world environment interactions vary with network conditions, provider load, rate limits, retries, and response drift, making repeatable comparisons across system configurations difficult. To support controlled comparisons, we combine actual model inference with recorded environment responses and service durations from historical agent trajectories. Replay runs within each comparison use the same initial messages, stage sequences, per-turn target output lengths, and recorded environment responses and durations.

\begin{figure}[!htbp]
\centering
\includegraphics[width=\linewidth]{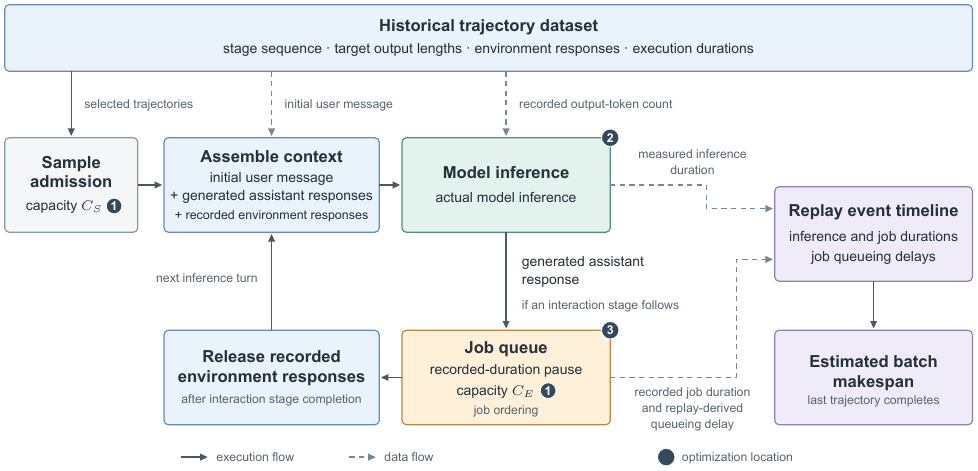}%
\caption{Overview of the controlled trace-driven agent replay framework. The numbered markers indicate optimization locations. \textcircled{\scriptsize 1}~Capacity separation: the sample-admission limit \(C_S\) is set independently of the environment-action concurrency limit \(C_E\). \textcircled{\scriptsize 2}~Inference optimizations: cache allocation, routing-prefix scheduling, and LBCH routing. \textcircled{\scriptsize 3}~CAAS for job ordering.}
\label{fig:trace-replay-framework}
\end{figure}

Figure~\ref{fig:trace-replay-framework} illustrates the turn-by-turn replay of a trajectory. At each inference turn, the input consists of the initial user message followed, in execution order, by model responses generated in preceding replay turns and recorded environment responses released earlier in the replay. A deployed SGLang service performs actual inference, generating the same number of output tokens as the corresponding historical turn. If an environment stage follows, the trajectory waits for an available execution slot and then pauses for the recorded job duration. The framework subsequently appends the recorded environment response to the context, making the next inference turn ready.

No environment actions execute during replay. Stage dependencies and environment durations remain fixed, while inference latency and cache reuse arise from actual serving. The replay timeline includes measured inference times, recorded environment durations, and queueing delays under the selected admission limits, execution capacity, and scheduling policies. Batch makespan is the elapsed time until all trajectories complete. Replay therefore measures execution efficiency rather than semantic correctness, as generated text need not reproduce the recorded action.

All replay experiments use Ling-3.0-Flash with thinking enabled, served by SGLang on H20 GPUs with four-way tensor parallelism (TP4) per worker. The model combines KDA and Multi-head Latent Attention layers at a 5:1 ratio with a sparse MoE backbone~\citep{ref50}. Inference-only replay covers SWE-bench Multilingual (SWE)~\citep{ref-swe-multilingual}, BFCL-v4 (BFCL)~\citep{ref54}, and WideSearch~\citep{ref60}, bypassing environment durations and capacity limits. Coordination replay uses MCP-Atlas~\citep{ref51} with recorded environment stages. Within each replay comparison, we keep the checkpoint, sampling parameters, workload, hardware budget, submission order, and target output lengths fixed. Appendix~\ref{app:inference-details} provides the inference-only replay workload details.\par

\begingroup
\clubpenalty=3000
\widowpenalty=3000
\subsection{Inference-Side Evaluation}
\label{sec:inference-evaluation}

\subsubsection{Cache allocation and admission operating point}
\label{sec:cache-allocation-evaluation}

We evaluate the joint effect of cache allocation and sample admission on a single TP4 worker with FCFS scheduling. Using Equation~\ref{eq:balanced-allocation-ratio}, we obtain \(r^{\mathrm{bal}}=0.09\) for SWE, \(r^{\mathrm{bal}}=2.26\) for BFCL, and \(r^{\mathrm{bal}}=0.10\) for WideSearch. The corresponding reference concurrency, \(C_{\mathrm{ref}}=\lfloor C(r^{\mathrm{bal}};\mathcal D)\rfloor\), is 98, 859, and 98, respectively. We perform a grid search over cache ratios and sample-concurrency limits around these model-derived reference points. Figure~\ref{fig:cache-allocation-sweep} summarizes the resulting makespans.

\textbf{Joint operating region.}
Under FCFS scheduling, the model-derived operating points yield makespans close to the best measured values across all three workloads, leaving only modest room for further improvement through local search. The analytical capacity model abstracts away allocation granularity and runtime fragmentation. Nevertheless, these results show that it provides a simple yet effective heuristic for jointly selecting the cache ratio and sample concurrency by accounting for trajectory token footprints and the number of trajectories that can remain resident.\par

\begin{figure}[!htbp]
\centering
\includegraphics[width=0.96\linewidth]{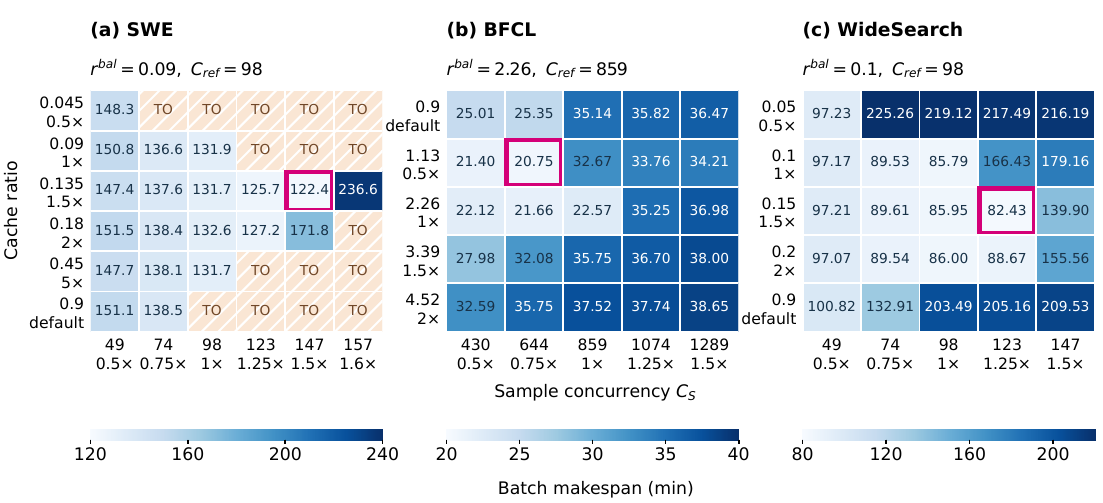}%
\caption{Cache allocation and sample-concurrency sweeps on (a) SWE, (b) BFCL, and (c) WideSearch. Cell values are complete-batch makespans in minutes; the panels use separate color scales. Second tick-label lines indicate multiples of \(r^{\mathrm{bal}}\) or \(C_{\mathrm{ref}}\), except for the default ratio. Magenta outlines mark the best measured points. TO denotes a six-hour timeout and is excluded from the continuous color scale.}
\label{fig:cache-allocation-sweep}
\end{figure}

The sweeps around the model-derived operating points show that the benefit of increasing concurrency depends on the cache ratio. At a fixed ratio, raising concurrency from below \(C_{\mathrm{ref}}\) can initially shorten makespan, but further increases may lead to sharp slowdowns or, on SWE, six-hour timeouts. Conversely, at a fixed concurrency, makespan can decrease and then increase as the ratio grows. Furthermore, at sufficiently low concurrency, makespan is relatively insensitive to the ratio, whereas ratios near \(r^{\mathrm{bal}}\) generally yield shorter makespans at higher concurrency. We next examine cache reuse to explain these trends.\par

\textbf{Cache reuse and the cost of excess concurrency.}
This limited sensitivity at low concurrency is consistent with both cache pools having sufficient capacity to retain reusable state across a broad range of ratios. At higher concurrency, an imbalanced partition can make one pool a bottleneck, whereas ratios near \(r^{\mathrm{bal}}\) balance the two residency capacities and help preserve cross-turn reuse. \par
\Needspace{4\baselineskip}
Figures~\ref{fig:cache-allocation-mechanism} and~\ref{fig:cache-service-throughput} relate the concurrency trends to cache reuse and system throughput at each workload's best measured cache ratio. Cache reuse measures the fraction of input tokens served from cache, while new-prefill work counts input tokens processed without cached reuse. We also report the time-averaged number of requests in service, including waiting, and output throughput, measured as successful output tokens per second.

\begin{figure}[!htbp]
\centering
\includegraphics[width=0.96\linewidth]{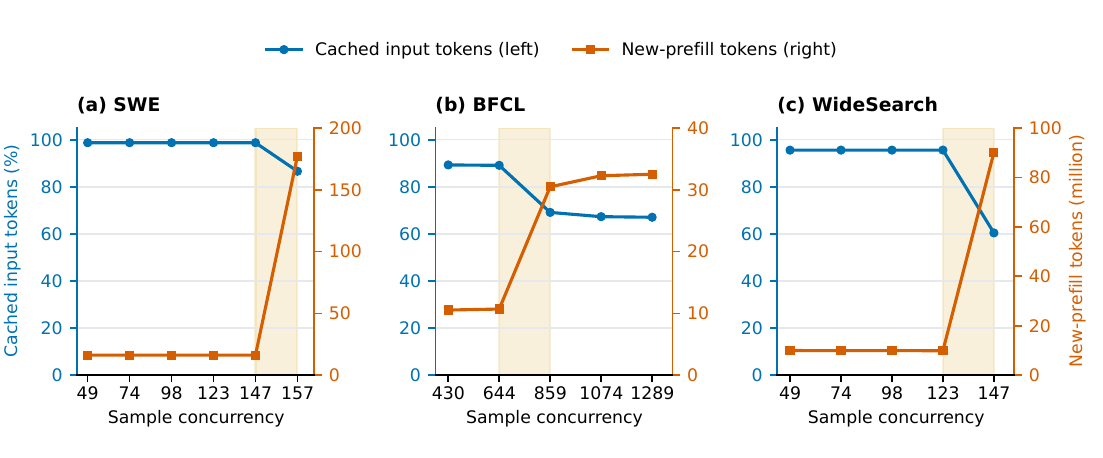}%
\caption{Cache reuse and cumulative new-prefill work versus sample concurrency at the best measured ratio for each workload: (a) SWE, \(r=0.135\); (b) BFCL, \(r=1.13\); (c) WideSearch, \(r=0.15\). Blue circles use the left axis and orange squares the right axis. Every point represents a complete batch. Concurrency configurations are equally spaced on the horizontal axis. Shading highlights the concurrency ranges where cache reuse drops sharply and prefill recomputation increases (SWE: 147--157; BFCL: 644--859; WideSearch: 123--147).}
\label{fig:cache-allocation-mechanism}
\end{figure}

Within the efficient operating region where cache reuse remains high and stable, increasing concurrency leaves new-prefill work largely unchanged while reducing makespan. The mean number of requests in service and output throughput increase together (Figure~\ref{fig:cache-service-throughput}). This pattern is consistent with higher concurrency improving GPU utilization through increased inference parallelism, thereby raising system throughput without substantial additional prefill work.\par

\begin{figure}[!htbp]
\centering
\includegraphics[width=0.96\linewidth]{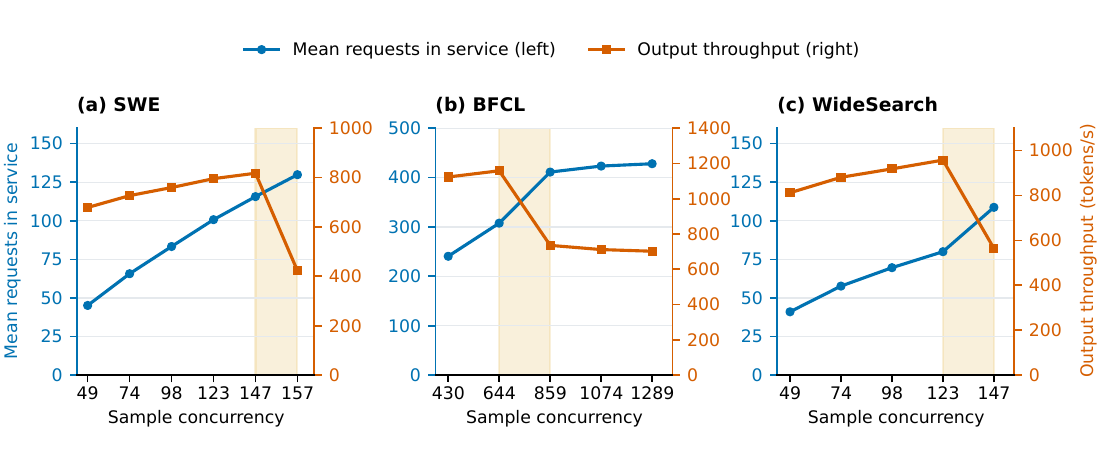}%
\caption{Mean requests in service (left) and output throughput (right) versus sample concurrency at the same cache ratios as Figure~\ref{fig:cache-allocation-mechanism}: (a) SWE; (b) BFCL; (c) WideSearch. Concurrency configurations are equally spaced on the horizontal axis. Shading highlights the concurrency ranges where output throughput drops despite increasing requests in service (SWE: 147--157; BFCL: 644--859; WideSearch: 123--147).}
\label{fig:cache-service-throughput}
\end{figure}

The shaded regions in Figures~\ref{fig:cache-allocation-mechanism} and~\ref{fig:cache-service-throughput} highlight where this relationship reverses. Increasing \(C_S\) from 147 to 157 for SWE, from 644 to 859 for BFCL, and from 123 to 147 for WideSearch causes cache reuse to decline and new-prefill work to rise to approximately eleven, three, and nine times its previous level, respectively. More requests reside in the engine, yet output throughput falls and makespan increases. Higher concurrency increases the demand for cached state from concurrently active trajectories. The resulting memory pressure can cause reusable cache entries to be evicted, reducing prefix reuse and increasing prefill recomputation.\par

Therefore, different cache ratios support different concurrency ranges for effective cross-turn cache reuse. Within these ranges, higher concurrency can improve inference parallelism and throughput; beyond them, prefix recomputation can outweigh the gains from additional parallelism. The cache ratio and sample concurrency should be selected jointly.\par

\subsubsection{In-worker scheduling}
\label{sec:in-worker-scheduling}

We compare FCFS, SGLang's native longest prefix match (LPM) and depth-first search weighting (DFS-weight), and routing-prefix scheduling on a single TP4 worker using the same inference-only workloads. We fix the cache ratio at the model-derived balanced allocation and vary sample concurrency to evaluate scheduling performance under low, moderate, and high cache pressure. Policy settings are provided in Appendix~\ref{app:inference-details}.

\textbf{Performance under cache pressure.}
Figure~\ref{fig:scheduling-makespan} compares complete-batch makespans. At low concurrency, makespans are generally close across scheduling policies, consistent with limited cache competition and little queueing, leaving relatively few opportunities to benefit from reordering requests. At the two highest tested concurrency levels, all three cache-aware policies achieve shorter makespans than FCFS across the three workloads. Together with the cache-reuse measurements discussed below, this contrast highlights the importance of identifying and prioritizing requests with reusable cached state to accelerate multi-turn agent inference under cache pressure. Among the three cache-aware policies, routing-prefix achieves shorter makespans than LPM and DFS-weight at most tested concurrency settings.

\begin{figure}[!htbp]
\centering
\includegraphics[width=0.90\linewidth]{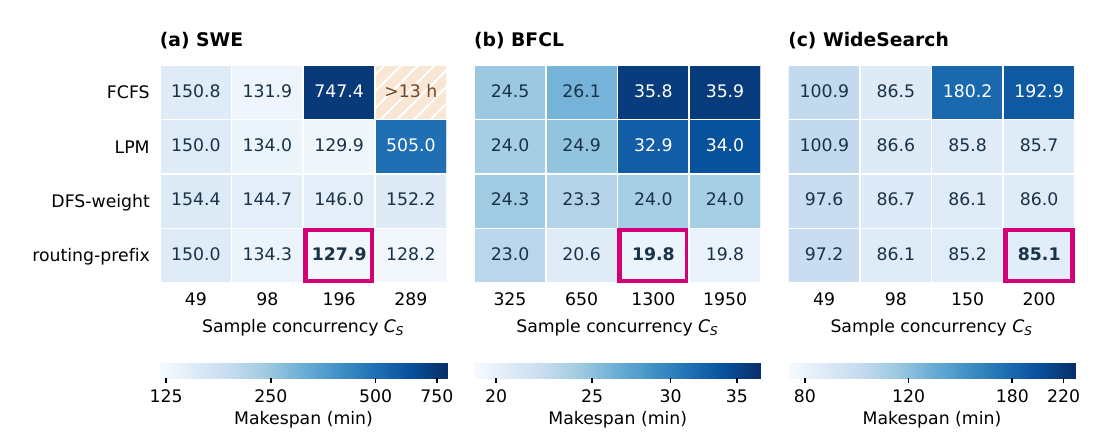}%
\caption{In-worker scheduling: (a) SWE; (b) BFCL; (c) WideSearch. Cell values are complete-batch makespans in minutes, with separate logarithmic color scales. Magenta outlines mark the minimum makespan for each workload before rounding. The hatched cell marked \(>13\,\mathrm{h}\) denotes a timeout for SWE FCFS at \(C_S=289\) under the 13-hour limit.}
\label{fig:scheduling-makespan}
\end{figure}

\textbf{Cache reuse and scheduling overhead.}
Figure~\ref{fig:scheduling-cache} relates these makespan differences to cache reuse and new-prefill work. At the lowest concurrency, the cache metrics are close across policies. As concurrency increases, routing-prefix continues to achieve high cache reuse with little additional prefill work. \par
\enlargethispage{\baselineskip}
\begin{figure}[H]
\centering
\includegraphics[width=0.88\linewidth]{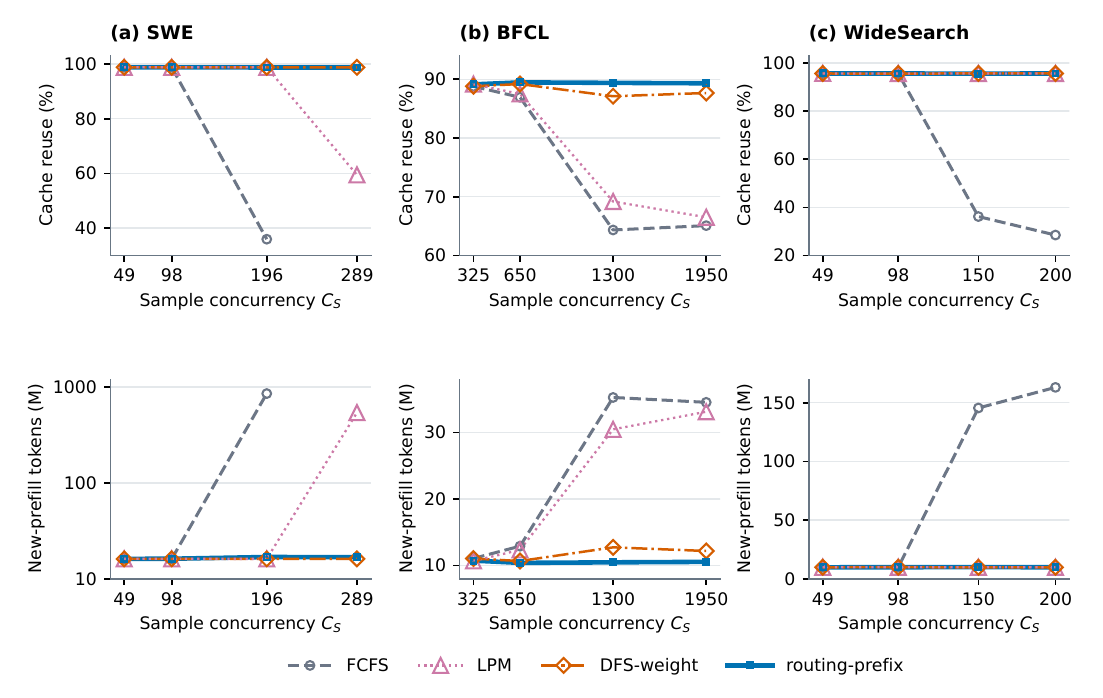}%
\caption{Cache reuse (top) and cumulative new-prefill work (bottom) from the scheduling runs under the configurations in Figure~\ref{fig:scheduling-makespan}. Each point represents a complete batch; the SWE FCFS timeout at \(C_S=289\) is omitted. The SWE new-prefill axis is logarithmic. The LPM, DFS-weight, and routing-prefix curves nearly overlap on WideSearch.}
\label{fig:scheduling-cache}
\end{figure}
\Needspace{4\baselineskip}
On SWE, routing-prefix and DFS-weight maintain similarly high cache reuse and low new-prefill work across the tested concurrency levels, whereas LPM shows a sharp drop in reuse and a corresponding rise in new-prefill tokens at the highest concurrency. On BFCL, routing-prefix achieves higher cache reuse and fewer new-prefill tokens than LPM and DFS-weight at the two highest concurrency levels, with a much greater advantage over LPM. On WideSearch, all three cache-aware policies maintain similarly high cache reuse and low new-prefill work, with their curves nearly overlapping. Together, these results highlight routing-prefix's consistently high cache reuse and low new-prefill work across the evaluated workloads and concurrency levels.\par
\enlargethispage{\baselineskip}

Cache reuse alone, however, does not fully explain why routing-prefix achieves shorter makespans even when cache metrics are similar across policies. We therefore examine scheduling overhead. Table~\ref{tab:scheduling-overhead} reports cumulative priority-calculation time at the highest tested effective sample concurrency for each workload. Routing-prefix incurs less overhead than DFS-weight on all three workloads and less than LPM on SWE and WideSearch. LPM repeatedly matches prefixes for queued requests when computing priorities, while DFS-weight additionally traverses the cache tree. Routing-prefix limits this work through selective prefix matching and reuse of recorded hit lengths. LPM has lower overhead on BFCL because it frequently falls back to FCFS when its waiting queue exceeds SGLang's built-in threshold of 128 requests. This fallback is intended to limit prefix-matching and sorting overhead on long queues, but bypasses cache-aware prioritization. Its lower calculation cost therefore does not by itself imply more efficient inference execution.

\begingroup
\setlength{\intextsep}{3pt}
\begin{table}[H]
\centering
\caption{Cumulative wall-clock time spent computing scheduling priorities over a complete batch at the highest tested effective sample concurrency for each workload. Bold values mark the minimum in each row.}
\label{tab:scheduling-overhead}
\begin{tabular}{@{}lr rrr@{}}
\toprule
Workload & \(C_S\) & LPM (s) & DFS-weight (s) & routing-prefix (s) \\
\midrule
SWE & 289 & 735.00 & 1704.25 & \textbf{303.05} \\
BFCL & 1950 & \textbf{7.65} & 172.15 & 10.91 \\
WideSearch & 200 & 62.82 & 144.08 & \textbf{39.97} \\
\bottomrule
\end{tabular}
\end{table}
\endgroup

By prioritizing queued requests from recently served trajectories, routing-prefix promotes reuse before their cached state is evicted. Trajectory-aware scheduling complements cache allocation and sample admission. An appropriate cache ratio and concurrency setting can preserve cross-turn reuse and achieve short makespans even with FCFS. Routing-prefix helps sustain reuse with low scheduling overhead as cache competition intensifies, making performance less sensitive to increased concurrency.\par

\subsubsection{Cross-worker routing}
\label{sec:cross-worker-routing-evaluation}

We compare SGLang's native cache-aware routing (CA), native consistent hashing (CH), and LBCH on two TP4 workers using the same inference-only workloads. All workers use routing-prefix scheduling, and the cache ratio and sample-concurrency limit are kept fixed across routing policies within each workload. The workload-specific configurations and policy parameters are listed in Appendix~\ref{app:inference-details}.

\textbf{Batch makespan and cache reuse.}
Figure~\ref{fig:routing-makespan} shows that LBCH achieves the shortest measured makespan on all three workloads, with reductions of 1.5\%--13.6\% relative to CA and 1.9\%--3.8\% relative to CH. Figure~\ref{fig:routing-mechanism} compares cache reuse and new-prefill work. Cache reuse rates are similar across policies within each workload (98.4\%--98.8\% on SWE, 86.2\%--87.2\% on BFCL, and 95.2\%--95.7\% on WideSearch), and LBCH does not consistently minimize new-prefill work.
Therefore, similar cache reuse does not necessarily translate into similar makespans, motivating a closer examination of how routing policies distribute requests across workers.

\begin{figure}[!htbp]
\centering
\includegraphics[width=0.96\linewidth]{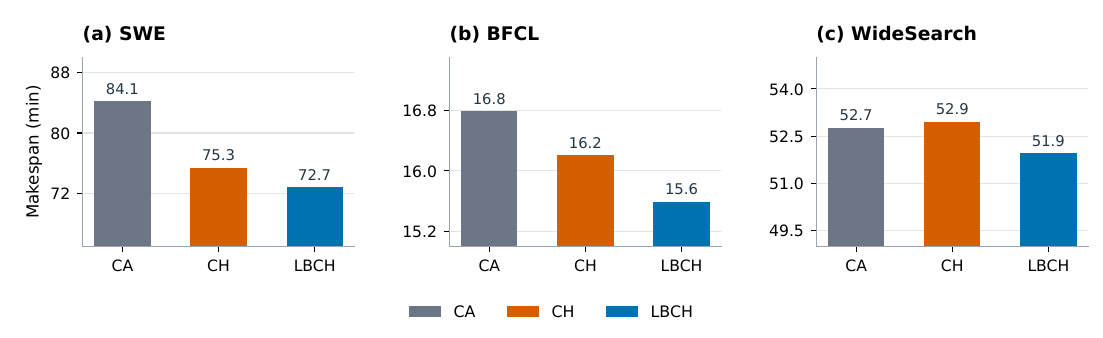}%
\caption{Cross-worker routing on two TP4 workers: (a) SWE; (b) BFCL; (c) WideSearch. Bars show measured batch makespan in minutes; the panels use separate vertical scales that do not start at zero.}
\label{fig:routing-makespan}
\end{figure}

\begin{figure}[!htbp]
\centering
\includegraphics[width=0.96\linewidth]{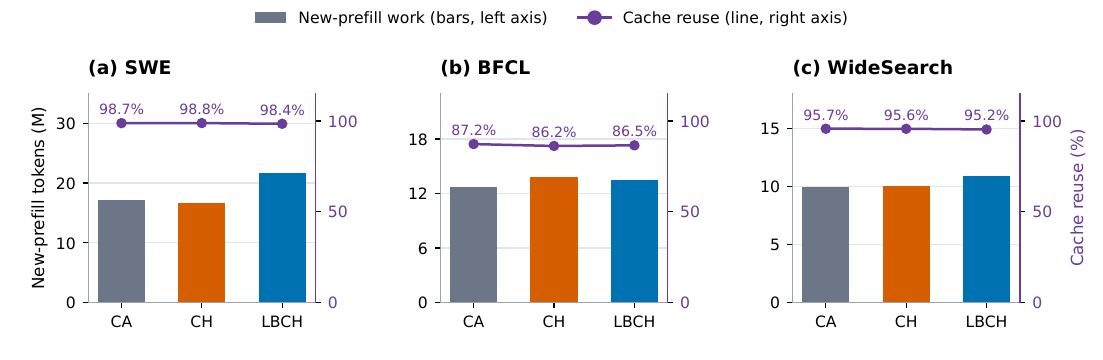}%
\caption{Cache reuse and new-prefill work for the runs in Figure~\ref{fig:routing-makespan}: (a) SWE; (b) BFCL; (c) WideSearch. Bars show new-prefill work on the left axis; the purple line shows cache reuse on the right axis. Workload panels use separate scales for new-prefill work.}
\label{fig:routing-mechanism}
\end{figure}

\begin{figure}[!htbp]
\centering
\includegraphics[width=0.96\linewidth]{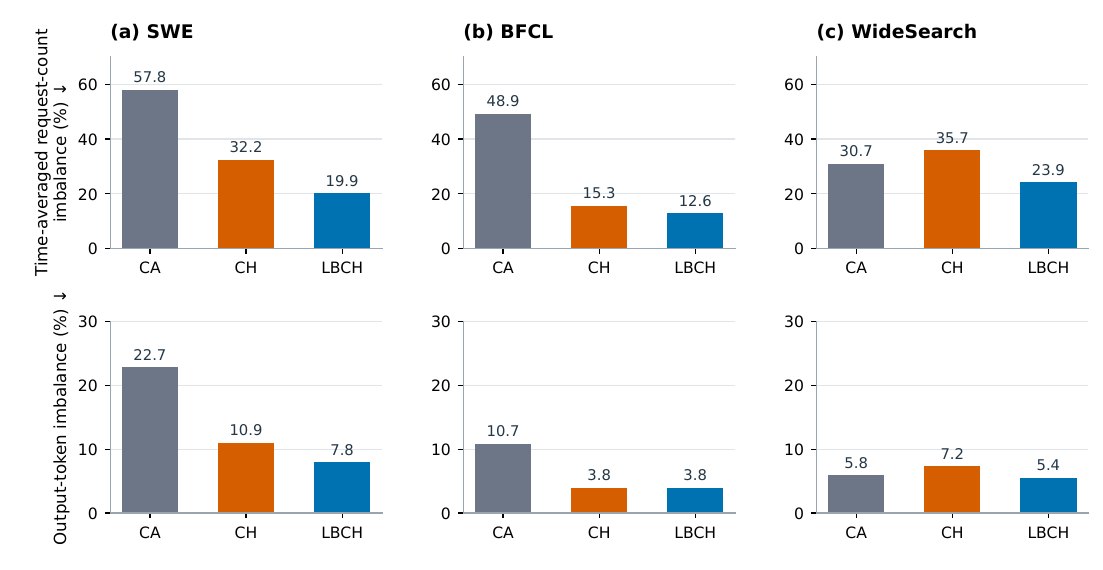}
\caption{Routing load distribution on (a) SWE, (b) BFCL, and (c) WideSearch. Top: time-averaged request-count imbalance over intervals with more than one in-flight request across both workers; each unit of retained time has equal weight. Bottom: output-token imbalance over the complete batch. Each row uses a common percentage scale. Lower values indicate more even distributions of requests over time and generated tokens across workers, respectively.}
\label{fig:routing-balance}
\end{figure}

\textbf{Load distribution.}
To assess load distribution across workers, we use two complementary metrics. Let \(N_w(t)\) denote the number of in-flight inference requests, including queued and executing requests, on worker \(w\) at time \(t\). \par
\Needspace{7\baselineskip}
Let \(T\) span from the start of the first request to the completion of the last request, and let \(J=\{t\in[0,T]:N_1(t)+N_2(t)>1\}\) denote the intervals with more than one in-flight request across both workers. We denote the time-averaged request-count imbalance over these retained intervals by \(\beta\), defined as
\begin{equation}
\beta=\frac{1}{|J|}\int_J d(t)\,dt,
\label{eq:routing-request-imbalance}
\end{equation}
where \(|J|\) is the total retained duration and \(d(t)\) is the instantaneous request-count imbalance,
\begin{equation}
d(t)=\frac{|N_1(t)-N_2(t)|}{N_1(t)+N_2(t)},\qquad t\in J.
\label{eq:routing-instantaneous-imbalance}
\end{equation}\par
This metric gives equal weight to each unit of retained elapsed time.
We also measure output-token imbalance, denoted by \(\gamma_{\mathrm{out}}\), to capture differences in generation workload, since request counts alone do not capture differences in output length. Let \(D_w\) be the total number of output tokens generated by requests executed on worker \(w\) over the batch:
\begin{equation}
D_w=\sum_{i\in\mathcal R_w}O_i,
\label{eq:routing-worker-output-tokens}
\end{equation}
where \(O_i\) is request \(i\)'s actual output length and \(\mathcal R_w\) contains requests actually executed on worker \(w\). We define \(\gamma_{\mathrm{out}}\) as
\begin{equation}
\gamma_{\mathrm{out}}=\frac{|D_1-D_2|}{D_1+D_2}.
\label{eq:routing-output-imbalance}
\end{equation}
A lower \(\beta\) indicates more balanced in-flight request counts across workers over time, while a lower \(\gamma_{\mathrm{out}}\) indicates a more balanced distribution of generated tokens across workers over the complete batch. Note that these metrics characterize load distribution rather than directly measuring computational cost.
Figure~\ref{fig:routing-balance} shows that LBCH achieves lower request-count imbalance than CA and CH across all three workloads. Its advantage over CA is particularly pronounced on SWE and BFCL, where output-token imbalance is also substantially lower. Compared with CH, LBCH improves both metrics on SWE and WideSearch, while on BFCL it achieves lower request-count imbalance with nearly identical output-token imbalance. Together with the cache-reuse results, these observations show that LBCH maintains high reuse while improving load balance, even though it does not always minimize new-prefill work. These results highlight the importance of jointly considering cache reuse and load distribution when routing multi-turn agent requests.

\par
\endgroup
\subsection{Inference--Environment Coordination}
\label{sec:tool-coordination-evaluation}

Using the replay framework in Section~\ref{sec:replay-framework}, we evaluate capacity separation and CAAS under constrained environment-execution capacity on historical trajectories from MCP-Atlas~\citep{ref51}, whose tasks rely on external MCP services that may degrade under high concurrency or encounter upstream rate limits. The inference deployment uses two TP4 workers with routing-prefix scheduling and LBCH. Guided by the capacity model in Section~\ref{sec:allocation}, we use \(C_S=50\) as the inference-side admission setting. To isolate the effect of capacity separation, we compare unified admission (\(C_S=C_E\)) with decoupled admission at \(C_S\in\{10,25,50\}\) for environment-execution limits of \(C_E\in\{3,5\}\), both with FCFS job scheduling. We then compare CAAS with FCFS at \(C_S=50\) under the same \(C_E\), changing only job ordering while keeping the inference configuration and recorded workload unchanged. Configuration details and supporting background on environment constraints are provided in Appendix~\ref{app:coordination-details}.

\begin{figure}[!htbp]
\centering
\includegraphics[width=0.72\linewidth]{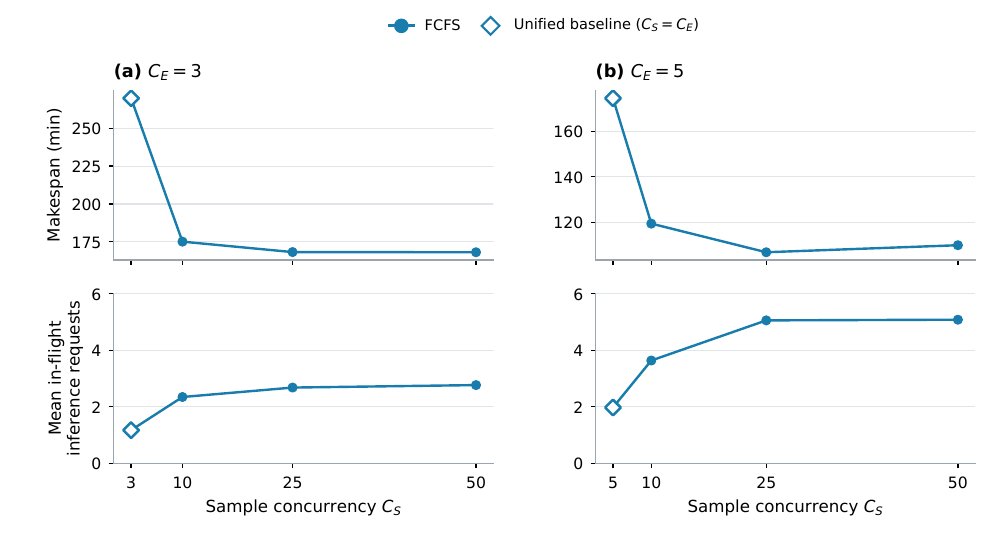}%
\caption{Capacity separation under FCFS job scheduling: batch makespan (top) and mean in-flight inference requests (bottom). The first point uses unified admission \(C_S=C_E\). In-flight requests are measured from client request launch to response completion and averaged over the replay makespan.}
\label{fig:coordination-capacity}
\end{figure}

\textbf{Capacity separation.}
Figure~\ref{fig:coordination-capacity} shows the effect of capacity separation at fixed environment-execution capacity. \par
\Needspace{3\baselineskip}
Increasing sample admission from the unified baseline to \(C_S=10\) reduces makespan by approximately one third at both environment-execution capacities. Capacity separation thus provides substantial gains even at modest sample concurrency. Further increases bring smaller improvements, with makespan fluctuating within a narrow range at \(C_S=25,50\).

The mean number of in-flight inference requests helps explain this pattern. It rises substantially even when sample admission is increased only to \(C_S=10\), but barely changes when \(C_S\) doubles from 25 to 50 at either environment-execution capacity. By supplying more concurrent inference work, capacity separation can improve GPU utilization and thereby increase overall system throughput. Admitting more samples allows inference and environment-interaction stages to overlap across trajectories, yet the limited environment-execution capacity constrains the release of subsequent inference work. Additional samples may then spend more time waiting for environment interaction rather than supplying concurrent inference work, so the makespan gains level off. At matched sample concurrency, \(C_E=5\) supports more in-flight requests and shorter makespans than \(C_E=3\), consistent with earlier release of subsequent inference work.\par

\begin{figure}[!htbp]
\centering
\includegraphics[width=0.82\linewidth]{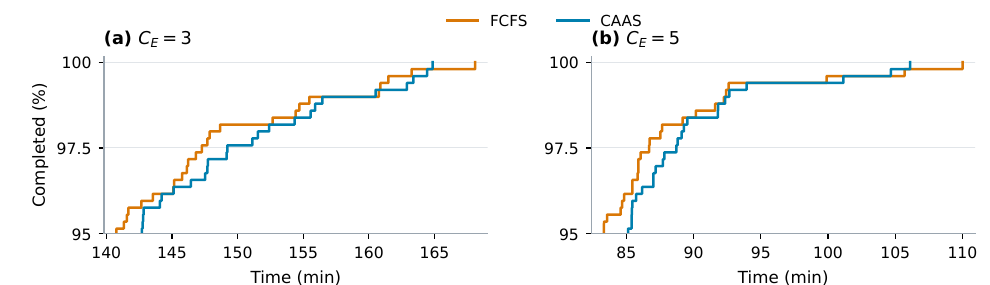}%
\caption{Completion of the final 5\% of trajectories at \(C_S=50\). Both policies use the same absolute replay time axis within each panel. The final-5\% tail duration is denoted by $L_5$.}
\label{fig:coordination-completion}
\end{figure}

\begin{samepage}
\textbf{CAAS and the completion tail.}
We evaluate CAAS at $C_S=50$ after capacity separation, where it reduces makespan relative to FCFS by 1.9\% at $C_E=3$ and 3.5\% at $C_E=5$. To examine how CAAS affects completion of the last few tasks, we measure the duration of the batch tail. Measuring time from the first task admission, let $t_q$ denote when the completed fraction first reaches or exceeds $q\%$, for $0<q\leq100$; $t_{100}$ is the batch makespan. The final-$p\%$ tail is the interval from $t_{100-p}$ to $t_{100}$, with duration
\begin{equation}
L_p=t_{100}-t_{100-p},\qquad 0<p<100.
\label{eq:batch-tail-duration}
\end{equation}
We use $p=5$ to analyze the replay completion tail below.\par
\end{samepage}
 Figure~\ref{fig:coordination-completion} shows that CAAS shortens the tail interval relative to FCFS by 19.0\% at $C_E=3$ and 21.4\% at $C_E=5$. Although CAAS reaches 95\% completion later, it finishes the batch earlier. Prioritizing jobs from trajectories with greater revealed depth can delay other trajectories, but allows subsequent stages of deeper chains to execute earlier, shortening the final completion tail. This behavior supports its design goal of exposing and advancing later stages of long execution chains before they become batch-ending bottlenecks, using only trajectory progress observed during the current run. These gains come from changing job dispatch order, not service times.

\subsection{End-to-End Validation}
\label{sec:end-to-end-validation}

We conduct live evaluations on BFCL, SWE, WideSearch, and MCP-Atlas using Ling-3.0-Flash (124B total and 5.1B active parameters)~\citep{ref50} and Ling-3.1-Flash (560B total and 25B active parameters)~\citep{ref-ling31}.
We omit BFCL-v4 for Ling-3.1-Flash, as scores are nearly saturated and we no longer use this benchmark in model development.
The two models use two TP4 workers on eight H20 GPUs and two TP8 workers on sixteen H20 GPUs, respectively.
Within each model--workload comparison, we fix the checkpoint, task set, sampling parameters, and hardware, varying only cache allocation, sample concurrency, inference scheduling and routing, and, for MCP-Atlas, capacity separation and environment-job scheduling.
Live makespan includes all samples and retry time, with no exclusion of samples that are retried or ultimately fail. For BFCL, SWE, and WideSearch, we tune sample concurrency for the original baselines with the default cache ratio, FCFS scheduling, and cache-aware routing fixed, targeting GPU saturation without substantial inference queueing.
The optimized configuration combines cache allocation and sample concurrency settings guided by the workload-aware capacity model with RP scheduling and LBCH routing.
For each model, SWE includes a Baseline-2 configuration that uses the same sample concurrency as the optimized configuration while retaining the baseline cache allocation and inference policies.
Detailed configurations and operating points are provided in Appendix~\ref{app:live-details} and Table~\ref{tab:live-configurations}.

\begin{table}[!htbp]
\centering
\caption{Live inference-side results. Each metric is calculated per run and averaged equally over three runs per configuration. Input token counts include cached tokens. Token totals are in millions. Imbalance metrics follow Section~\ref{sec:cross-worker-routing-evaluation}.}
\label{tab:live-end-to-end}
\small
\setlength{\tabcolsep}{4pt}
\begin{tabular*}{\linewidth}{@{\extracolsep{\fill}}lllrrrrrrr@{}}
\toprule
 & & & Makespan & \multicolumn{2}{c}{Tokens (M)} & Reuse & \multicolumn{2}{c}{Imbalance (\%)} & \\
\cmidrule(lr){5-6}\cmidrule(lr){8-9}
Model & Workload & Config. & (min) & Input & Output & \(H\) (\%) & \(\beta\) & \(\gamma_{\mathrm{out}}\) & Score \\
\midrule
3.0 & BFCL       & Baseline   & 68.4  & 141.1  & 3.9  & 84.2 & 10.6 & 2.9  & 73.0 \\
    &            & Opt        & 49.2  & 141.6  & 3.9  & 84.5 & 17.0 & 3.1  & 75.65 \\
    & SWE        & Baseline-1 & 166.9 & 1286.2 & 8.5  & 98.3 & 17.2 & 13.0 & 72.4 \\
    &            & Baseline-2 & 197.7 & 1371.4 & 8.7  & 87.5 & 17.2 & 7.7  & 71.3 \\
    &            & Opt        & 149.5 & 1347.9 & 8.7  & 98.6 & 16.4 & 3.5  & 72.6 \\
    & WideSearch & Baseline   & 186.7 & 261.9  & 10.9 & 79.9 & 11.5 & 1.4  & 73.6 \\
    &            & Opt        & 116.7 & 311.1  & 10.0 & 81.7 & 15.6 & 4.6  & 72.7 \\
\midrule
3.1 & SWE        & Baseline-1 & 234.1 & 1073.3 & 11.1 & 98.2 & 15.1 & 11.6 & 87.16 \\
    &            & Baseline-2 & 304.1 & 1102.8 & 11.5 & 80.4 & 15.9 & 12.0 & 86.0 \\
    &            & Opt        & 185.1 & 1045.7 & 11.0 & 98.3 & 16.2 & 3.3  & 88.33 \\
    & WideSearch & Baseline   & 459.9 & 415.7  & 25.0 & 83.2 & 8.8  & 1.7  & 78.6 \\
    &            & Opt        & 359.1 & 430.0  & 25.1 & 80.7 & 17.6 & 2.5  & 78.1 \\
\bottomrule
\end{tabular*}
\end{table}

\pagebreak
Table~\ref{tab:live-end-to-end} reports mean makespan, token counts, cache reuse, worker-load imbalance, and evaluation scores over three runs per configuration for the first three workloads, where environment-execution capacity is sufficient and does not constrain sample admission.
For Ling-3.0-Flash, the optimized configuration reduces mean makespan by 28.1\% on BFCL, 10.4\% on SWE, and 37.5\% on WideSearch relative to their original baselines.
For Ling-3.1-Flash, the corresponding reductions are 20.9\% on SWE and 21.9\% on WideSearch.
These results demonstrate that the proposed optimizations scale effectively to larger models while still achieving substantial makespan reductions.
Stochastic model sampling and environment variability cause trajectories and token counts to differ across live runs.
The makespan reductions, even with higher input-token volumes on some workloads, support the effectiveness of the optimizations under realistic execution variability.

On BFCL and WideSearch, optimized runs achieve shorter makespans despite higher worker-load imbalance, while maintaining comparable cache reuse.
These results suggest a more effective trade-off among concurrency, cache reuse, and worker-load balance, where the benefits of sustaining cache reuse at higher concurrency can outweigh a slight deterioration in worker-load balance.
The SWE comparison further shows that increased concurrency alone does not explain these gains.
For Ling-3.0-Flash, cache reuse falls from 98.3\% in Baseline-1 to 87.5\% in the higher-concurrency Baseline-2.
This decline is consistent with increased cache pressure and eviction, leading to additional recomputation that can offset parallelism gains.
At the same concurrency as Baseline-2, however, the optimized configuration maintains 98.6\% reuse and reduces makespan by 24.4\%.
Ling-3.1-Flash shows a similar pattern: with the same sample concurrency, Opt maintains 98.3\% cache reuse versus Baseline-2's 80.4\% and reduces makespan by 39.1\%.
Consistent with the controlled experiments in Sections~\ref{sec:cache-allocation-evaluation}--\ref{sec:cross-worker-routing-evaluation}, these results demonstrate the benefits of coordinated cache allocation, scheduling, and routing in preserving cache reuse at higher concurrency and reducing batch makespan. The latency measurements further clarify how these batch-level gains relate to individual request performance. Compared with the concurrency-matched Baseline-2, Opt reduces TTFT by 24.9\% and 50.2\%, and TPOT by 51.1\% and 40.2\%, for Ling-3.0-Flash and Ling-3.1-Flash, respectively. Relative to Baseline-1, Opt uses higher concurrency, increasing competition for inference resources; this is consistent with higher per-request TTFT and TPOT despite shorter batch makespan, reflecting the distinction between batch efficiency and per-request latency. Appendix~\ref{app:live-details} provides a detailed analysis of the SWE results.

\begin{table}[!htbp]
\centering
\caption{Live MCP-Atlas results, averaged equally over three 500-task runs per configuration. Tail durations $L_5$ and $L_{10}$ are measured from each run's own 95\% and 90\% completion points to its final task termination, including both successful and failed tasks. Mean in-flight is the time average of recorded inference calls over the full task-execution window, computed from native trajectory start/end timestamps and averaged equally across runs.}
\label{tab:live-mcp-tail}
\small
\begin{tabular*}{\linewidth}{@{\extracolsep{\fill}}llrrrrr@{}}
\toprule
Model & Configuration & Makespan (min) & \shortstack{Mean\\in-flight} & \(L_5\) (min) & \(L_{10}\) (min) & Score \\
\midrule
3.0 & Baseline & 466.1 & 1.2 & 38.9 & 73.7 & 65.5 \\
    & Opt-FCFS & 168.2 & 7.3 & 67.6 & 79.2 & 66.6 \\
    & Opt-CAAS & 163.0 & 7.9 & 66.0 & 73.4 & 65.6 \\
\midrule
3.1 & Baseline & 333.5 & 1.6 & 26.1 & 43.5 & 81.0 \\
    & Opt-FCFS & 140.5 & 12.5 & 49.8 & 54.4 & 80.9 \\
    & Opt-CAAS & 128.5 & 11.5 & 35.3 & 38.6 & 81.3 \\
\bottomrule
\end{tabular*}
\end{table}

For MCP-Atlas, Table~\ref{tab:live-mcp-tail} reports mean results over three 500-task runs per configuration, evaluating the integrated inference and environment optimizations under constrained environment capacity. Relative to the unified-admission baseline, inference-side optimizations and capacity separation (Opt-FCFS) reduce mean makespan from 466.1 to 168.2 minutes (63.9\%) for Ling-3.0-Flash and from 333.5 to 140.5 minutes (57.9\%) for Ling-3.1-Flash, with environment capacity unchanged for each model. Mean in-flight inference calls increase from 1.2 to 7.3 for Ling-3.0-Flash and from 1.6 to 12.5 for Ling-3.1-Flash, consistent with more inference work being available while other trajectories await environment interaction. \par

Compared with Opt-FCFS, Opt-CAAS further reduces mean makespan by 3.1\% for Ling-3.0-Flash and 8.5\% for Ling-3.1-Flash. Using the tail durations defined in Equation~\ref{eq:batch-tail-duration}, the corresponding reductions in $L_5$ and $L_{10}$ are 2.4\% and 7.3\% for Ling-3.0-Flash, and 29.1\% and 29.0\% for Ling-3.1-Flash. These live results are consistent with the benefits of CAAS observed in the controlled replay experiments in Section~\ref{sec:tool-coordination-evaluation}, although live trajectories vary across runs.

Although Opt-CAAS completes the batch much earlier, its final 5\% takes longer than the baseline's. This apparent discrepancy can be explained by differences in the tasks that finish last. Under limited admission, the baseline's final tasks are largely those admitted last, rather than necessarily those with the longest trajectories. Broader admission lets shorter trajectories finish while longer ones continue, so the optimized tail is more concentrated in long trajectories. The final 25 tasks under Opt-CAAS average 27.9 and 15.6 turns for Ling-3.0-Flash and Ling-3.1-Flash, respectively, compared with 10.1 and 7.9 under the baseline. When matched by task ID, the Opt-CAAS tail tasks also require more turns in baseline runs. This difference helps explain why Opt-CAAS can finish the batch much earlier while taking longer to complete its final 5\%, so $L_5$ alone should not be used to compare overall execution efficiency across different admission policies.

Across the evaluated model--benchmark combinations, evaluation scores remain aligned between the baseline and optimized configurations, indicating that the efficiency gains do not come at the expense of response quality.

\section{Discussion and Future Work}
\label{sec:discussion}

Although this work focuses on batch makespan in LLM training and development workflows, its underlying principles may extend to other agent workloads. For example, in personal-agent services, workload-aware cache allocation, RP scheduling, and LBCH routing could preserve cross-turn reuse while supporting concurrent users. These mechanisms could reduce prefix recomputation and shorten the prefill component of TTFT. Reducing redundant prefill may also help reduce TPOT by leaving more capacity for ongoing decoding. Separating sample admission from environment-execution capacity could let ready trajectories use GPUs while others wait for tools or external services. In workflows with parallel subagents, the chain-advancing principle behind CAAS could inform tool-job scheduling to advance subagents that delay downstream aggregation or overall task completion.

Many directions remain for improving agent trajectory execution efficiency and further reducing makespan. These include extending workload-aware cache allocation and trajectory-aware scheduling to hierarchical caching while accounting for cache capacity and transfer costs, and investigating cache eviction policies for hybrid models in agent workloads. Another important direction is makespan optimization subject to application service-level objectives (SLOs), such as TTFT and TPOT targets, which is relevant to datacenters seeking to improve task throughput while meeting user-facing latency requirements.

\Needspace{13\baselineskip}
\section{Conclusion}
\label{sec:conclusions}

We present coordinated inference and environment optimizations to reduce batch makespan for multi-turn agents served by hybrid LLMs. A workload-aware capacity model guides cache allocation and sample concurrency, while RP and LBCH promote timely cross-turn reuse at low scheduling cost and balance trajectory affinity with worker load. Under constrained environment resources, capacity separation improves inference--environment overlap, and CAAS advances long execution chains to shorten the batch tail. Controlled replay evaluates individual mechanisms, while live experiments with Ling-3.0-Flash and Ling-3.1-Flash demonstrate batch makespan reductions of 10.4\%--65.0\% across four well-established benchmarks without compromising model performance. Their successful application in developing the Ling-3.x model series demonstrates the practical value of these optimizations in improving model-development efficiency.

\begingroup
\renewcommand{\bibfont}{\fontsize{10}{13}\selectfont\raggedright}
\setlength{\bibsep}{6pt}

\endgroup

\clearpage
\appendix
\fontsize{10}{12}\selectfont
\setlength{\parskip}{3pt plus 1pt minus 1pt}
\setlength{\textfloatsep}{10pt plus 2pt minus 2pt}
\setlength{\intextsep}{8pt plus 2pt minus 2pt}
\setcounter{figure}{0}
\setcounter{table}{0}
\renewcommand{\thefigure}{A\arabic{figure}}
\renewcommand{\thetable}{A\arabic{table}}
\renewcommand{\theHfigure}{appendix.\arabic{figure}}
\renewcommand{\theHtable}{appendix.\arabic{table}}
\section{Experimental Configuration Details}
\label{app:experimental-details}

\subsection{Inference-Side Experiments}
\label{app:inference-details}

Table~\ref{tab:replay-workloads} lists the inference-only workloads. The SWE replay set contains 289 trajectories because only successful historical executions were collected; failed executions were excluded.
Selected SWE trajectories have long contexts and deep chains; BFCL lengths vary substantially and include single-turn trajectories to evaluate our methods on short trajectories; WideSearch repeatedly invokes inference for broad information seeking~\citep{ref-swe-multilingual,ref54,ref60}. Replay measures execution efficiency for these recorded workloads; it does not evaluate semantic correctness or cover all failures and retries.

\begin{table}[!htbp]
\centering
\caption{Historical inference-only replay workloads. Output totals are from the routing replay runs. Live task counts are reported separately in Table~\ref{tab:live-configurations}.}
\label{tab:replay-workloads}
\normalsize
\begin{tabular}{@{}lrrr@{}}
\toprule
Workload & Recorded trajectories & Inference requests & Output tokens \\
\midrule
SWE & 289 & 25,416 & 6,000,784 \\
BFCL & 5,106 & 17,029 & 1,441,547 \\
WideSearch & 200 & 5,672 & 4,736,147 \\
\bottomrule
\end{tabular}
\end{table}

\textbf{Cache allocation and admission.} The experiments in Section~\ref{sec:cache-allocation-evaluation} use one TP4 worker with FCFS scheduling. The cache is flushed before each timed run, and complete-batch makespan excludes deployment and warmup. The tested cache ratios and sample-concurrency limits are shown in Figure~\ref{fig:cache-allocation-sweep}; TO denotes a six-hour timeout.

\textbf{In-worker scheduling.} In Section~\ref{sec:in-worker-scheduling}, the cache ratio is fixed at the model-derived balanced allocation: 0.09 for SWE, 2.26 for BFCL, and 0.10 for WideSearch. LPM and DFS-weight use SGLang's default policy parameters. Figure~\ref{fig:scheduling-makespan} reports the tested sample-concurrency limits; the SWE FCFS run at the largest effective concurrency of 289 is marked as a timeout under the 13-hour limit.

\textbf{Cross-worker routing.} The replay comparison in Section~\ref{sec:cross-worker-routing-evaluation} uses two TP4 workers with routing-prefix scheduling. The fixed \((r,C_S)\) values are \((0.09,196)\) for SWE, \((2.26,2000)\) for BFCL, and \((0.10,200)\) for WideSearch. CA and CH use their native default parameters, and LBCH uses \(\alpha=1.25\). Request imbalance is averaged over the intervals with more than one in-flight request across both workers, as defined in Equation~\ref{eq:routing-request-imbalance}. Output imbalance uses the generated token totals assigned to each worker.

\subsection{Inference--Environment Coordination}
\label{app:coordination-details}

\textbf{Environment-capacity settings.} MCP-Atlas \citep{ref51} tasks depend on external MCP services, which may degrade under high concurrency or encounter upstream rate limits such as HTTP 429 errors. The runner configuration referenced for these experiments defaulted to five concurrent tasks.\footnote{See the official \href{https://github.com/scaleapi/mcp-atlas/blob/main/README.md}{MCP-Atlas runner configuration}, \href{https://github.com/scaleapi/mcp-atlas/blob/main/services/agent-environment/README.md}{tool-serving guidance}, and \href{https://docs.slack.dev/apis/web-api/rate-limits/}{Slack API rate limits}.} This motivates evaluating constrained environment-execution settings, rather than assuming that environment capacity grows with inference capacity. We report results at \(C_E\in\{3,5\}\).

\textbf{Replay configuration.} Section~\ref{sec:tool-coordination-evaluation} uses 495 successful historical trajectories with recorded environment responses and job service times. Two TP4 workers use a cache ratio of \(r=0.14\), routing-prefix scheduling, and LBCH routing. CAAS uses a time-band width of \(B=30\) seconds and an inference-duration bucket width of \(\Delta_I=1\) second. Compared replay runs keep the model, hardware, and recorded workload fixed.\par

\subsection{Live Execution Settings}
\label{app:live-details}

Both models enable thinking mode and use a maximum context length of 256K tokens in the live experiments. Table~\ref{tab:live-configurations} summarizes the task sets and operating points.

\begin{table}[!htbp]
\centering
\caption{Live task sets and operating points. All baseline cache ratios are 0.9. B1 and B2 denote SWE Baseline-1 and Baseline-2. The two MCP-Atlas Opt variants share the listed settings. A dash in \(C_E\) indicates that environment execution does not constrain sample admission.}
\label{tab:live-configurations}
\small
\begin{tabular}{@{}llrrrrrr@{}}
\toprule
 & & & \multicolumn{3}{c}{Sample concurrency \(C_S\)} & & \\
\cmidrule(lr){4-6}
Model & Workload & Tasks & Baseline/B1 & B2 & Opt & Opt ratio & \(C_E\) \\
\midrule
3.0 & BFCL       & 5,106 & 100 & --  & 300 & 1.5  & -- \\
    & SWE        & 300   & 100 & 250 & 250 & 0.1  & -- \\
    & WideSearch & 200   & 40  & --  & 100 & 0.25 & -- \\
    & MCP-Atlas  & 500   & 3   & --  & 50  & 0.14 & 3 \\
\midrule
3.1 & SWE        & 300   & 50  & 100 & 100 & 0.2  & -- \\
    & WideSearch & 200   & 15  & --  & 40  & 0.27 & -- \\
    & MCP-Atlas  & 500   & 3   & --  & 50  & 0.34 & 3 \\
\bottomrule
\end{tabular}
\end{table}

For MCP-Atlas, the baseline uses unified admission with $C_S=C_E=3$. Both Opt variants use the same inference configuration and decoupled admission at $C_S=50$ while retaining $C_E=3$. They differ only in FCFS versus CAAS environment dispatch.

For the live evaluations in Section~\ref{sec:end-to-end-validation}, to use the same load-balance criteria for CA and LBCH, CA sets \texttt{balance\_abs\_threshold}=0 and \texttt{balance\_rel\_threshold}=1.6666667, while LBCH uses $\alpha=1.25$. Live CAAS uses the same parameters as the replay experiments: $B=30$ seconds and $\Delta_I=1$ second.

Live WideSearch also uses the evaluated model for summarization, which can introduce additional uncached input and help explain its lower aggregate cache reuse relative to replay.

\textbf{Additional SWE diagnostics.} Table~\ref{tab:live-swe-detailed} provides a closer comparison of the tuned baseline, concurrency-matched baseline, and Opt.

\begin{table}[!htb]
\centering
\caption{Live SWE diagnostics under the tuned baseline, concurrency-matched baseline, and Opt. New-prefill counts uncached input tokens. TTFT and TPOT are averaged across requests within each run, then across runs. Mean API in-flight averages outstanding API requests, including those waiting, over each live run.}
\label{tab:live-swe-detailed}
\small
\setlength{\tabcolsep}{4pt}
\begin{tabular}{@{}lrrrrrr@{}}
\toprule
 & \multicolumn{3}{c}{Ling-3.0-Flash} & \multicolumn{3}{c}{Ling-3.1-Flash}\\
\cmidrule(lr){2-4}\cmidrule(lr){5-7}
Metric & Baseline-1 & Baseline-2 & Opt & Baseline-1 & Baseline-2 & Opt \\
\midrule
Sample concurrency & 100 & 250 & 250 & 50 & 100 & 100 \\
Makespan (min) & 166.9 & 197.7 & 149.5 & 234.1 & 304.1 & 185.1 \\
Cache reuse (\%) & 98.3 & 87.46 & 98.58 & 98.2 & 80.41 & 98.27 \\
New-prefill (M) & 22.2 & 170.90 & 18.98 & 19.3 & 211.81 & 17.93 \\
Mean API in-flight & 40.9 & 105.2 & 76.7 & 35.5 & 71.0 & 63.7 \\
TTFT (s) & 2.90 & 10.08 & 7.57 & 2.54 & 11.81 & 5.88 \\
TPOT (ms/token) & 28.6 & 85.55 & 41.81 & 24.3 & 60.84 & 36.37 \\
\bottomrule
\end{tabular}
\end{table}

We also include Baseline-LowConcurrency as a reference configuration. Baseline-LowConcurrency uses $C_S=20$ with the default cache ratio, FCFS scheduling, and cache-aware routing. Its live SWE runs yield mean makespans of 208.1 and 263.0 minutes for Ling-3.0-Flash and Ling-3.1-Flash, respectively, over three runs per configuration. Tuning sample concurrency under this baseline configuration yields Baseline-1 at $C_S=100$ and 50, respectively, reducing makespan to 166.9 and 234.1 minutes (19.8\% and 11.0\%) while maintaining cache reuse above 98\%. This is consistent with the concurrency benefits observed in Section~\ref{sec:cache-allocation-evaluation}.

Increasing sample concurrency from Baseline-1 to Baseline-2 lowers cache reuse and increases new-prefill work, request latency, and makespan for both models. At the same sample concurrency as Baseline-2, Opt restores cache reuse to above 98\% and reduces new-prefill tokens by 88.9\% and 91.5\%, respectively. TTFT falls by 24.9\% and 50.2\%, and TPOT falls by 51.1\% and 40.2\%. Together with makespan reductions of 24.4\% and 39.1\%, these results show that Opt maintains high cache reuse at high concurrency, benefiting from both parallelism and cache reuse to shorten makespan.

At the same sample concurrency as Baseline-2, Opt has fewer mean in-flight requests, lower TTFT and TPOT, and shorter makespan. Together, these observations are consistent with faster request completion and task progress. Relative to Baseline-1, Opt uses higher concurrency, increasing competition for inference resources; this is consistent with higher per-request TTFT and TPOT despite shorter batch makespan, distinguishing batch efficiency from per-request latency.

\end{document}